\documentclass[journal,article,submit,pdftex,moreauthors]{Definitions/mdpi} 

\renewcommand{\linenumbers}{}
\firstpage{1} 
\pubvolume{1}
\issuenum{1}
\articlenumber{0}
\pubyear{2025}
\copyrightyear{2025}
\datereceived{ } 
\daterevised{ } 
\dateaccepted{ } 
\datepublished{ } 
\hreflink{https://doi.org/} 

\Title{Multiplicity Distributions and the Frontier between Soft and Hard Physics}

\TitleCitation{Multiplicity Distributions and the Frontier between Soft and Hard Physics} 

\Author{H. R. Martins-Fontes  and F. S. Navarra *}

\AuthorNames{H.R. Martins-Fontes$^{1}$  and F. S. Navarra$^{2,}$* }

\isAPAStyle{%
       \AuthorCitation{Martins-Fontes, H., Lastname, F., \& Lastname, F.}
         }{%
        \isChicagoStyle{%
        \AuthorCitation{Lastname, Firstname, Firstname Lastname, and Firstname Lastname.}
        }{
        \AuthorCitation{Lastname, F.; Lastname, F.; Lastname, F.}
        }
}

\address{%
$^{1}$ \quad Instituto de F\'{\i}sica - Universidade de S\~ao Paulo, Rua do Mat\~ao 1371, CEP 05508-090, S\~ao Paulo - Brasil; 
             henriquemartinsfontes@usp.br  \\
$^{2}$ \quad Instituto de F\'{\i}sica - Universidade de S\~ao Paulo, Rua do Mat\~ao 1371, CEP 05508-090, S\~ao Paulo - Brasil; 
            navarra@if.usp.br}

\corres{Correspondence: navarra@if.usp.br}

\secondnote{These authors contributed equally to this work.}

\abstract{The multiplicity distributions measured in proton proton collisions at the LHC exhibit interesting new features. One of them is
the appearance of substructures, such as the so-called "shoulder" at large multiplicities. The most natural interpretation of this behavior is the existence of two production mechanisms. The final result is then a superposition of two distributions. In a previous publication we assumed that the two production mechanisms 
are soft and semihard partonics scatterings.  In this work we further discuss this assumption and, in particular, 
we  study the dependence of the results on the scale which separates soft from hard events. 
}


\begin{document}



\section[Introduction]{Introduction}

Particle production at high energies is a complex phenomenon. In principle it might be completely understood with Quantum Chromodynamics
(QCD) but we are still far to reach this goal. QCD is well under control at high momentum transfer (of several GeV) reactions but it is
poorly understood at low (less than one GeV) momentum transfer. In the first case, the theory is in the perturbative regime (pQCD) and we can perform reliable calculations, which have been very successful in reproducing experimental data. 

It is expected that, at increasing energies we  progressively enter 
the perturbative domain and at asymptotically high energies all 
processes will be explained almost exclusively by pQCD. The question is: when does pQCD starts to be dominant ? 

Usually one applies pQCD to events where a hard energy scale (of the order of dozens of GeVs) is present, such as jet events. However, 
in \cite{gribov} Gribov, Levin and Ryskin  explored parton-parton scatterings with a momentum transfer of a few GeV and realized 
that  pQCD calculations were reliable even in that regime, which they called "semihard".  A few years later, Mueller and Navelet 
\cite{mueller} introduced the word "minijet" to describe the same kind of (semihard) process and usually we take these two terms as synonymous. 
Fourty years ago, Gaisser and Halzen \cite{halzen1} proposed that in proton proton collisions in the TeV region we would see clear 
effects of pQCD. They built a two-component model, in which the cross section was the sum of a soft and a semihard part. The former was 
assumed to be energy independent and the latter was calculated with pQCD. They assumed that minimum bias proton-proton collisions could 
be well described by  strongly energy dependent pQCD minijets supplemented by a weakly energy dependent soft component.  In the late eighties the accelerators approached the TeV region and some 
changes started to be seen in global inclusive observables, such as transverse momentum and multiplicity distributions. For Gaisser and Halzen these were manifestations of the onset of the QCD perturbative dynamics. The first results were promising. In a subsequent work   
\cite{halzen2}, this two-component model was able to reproduce the measured multiplicity distributions, giving a natural explanation to their double-hump structure. In these calculations there were integrals over the transverse momentum of the 
produced partons which were infrared divergent and needed to be cut at a minimum value, $p_{T min}$. Apart from regularizing the 
integrals,  this infrared cutoff defines the frontier between soft (below $p_{T min}$) and semihard (above $p_{T min}$) physics. 
In \cite{halzen1} the authors used $1 < p_{T min} < 2$ GeV for collisions at $\sqrt{s} \approx 1$ TeV.  This kind of minijet model
was further developed in \cite{sarcevic}, where  $p_{T min} \approx 3$ GeV was found to lead to the best agreement with data. Inspite of its
initial success the  two-component models based on additive cross sections, i.e. $\sigma_{tot} = \sigma_{soft} + \sigma_{semihard}$, 
and on the infrared cutoff $p_{T min}$ had two serious problems: first, the dependence of the cross sections on $p_{T min}$ was too 
strong and second, the cross sections would grow too fast with the energy, violating the unitarity bounds \cite{pi}.  
These difficulties were circumvented by the eikonalized minijet models (see \cite{luna} and references therein).

About twenty years later a new solution for these problems appeared. The advance of the studies of the quantum evolution of parton 
scattering amplitudes led to 
the theory of the Color Glass Condensate (CGC) \cite{cgc}. In the CGC, an intrinsic energy scale of the theory, the saturation scale 
$Q_s$, emerges and at the same times regularizes the theory and tames the growth of the parton distributions with the energy. However this formalism was not able to explain some of the most inclusive observables, such as the 
total proton-proton cross section, which, even
at the highest energies, seems to contain non-perturbative physics (not directly included in the CGC). Therefore the development of 
multi-component models remained relevant. In \cite{babi} a three component model was proposed to describe the dependence of the 
proton-proton total cross section with the energy.  Three energy ranges were defined: $0 < p_T <\Lambda_{QCD}$; 
$\Lambda_{QCD} <  p_T < Q_s$ and $Q_s < p_T$. It is important to notice that for most cases of practical interest the saturation scale 
is small ($< 1$ GeV) and can not play the role of separation scale, marking the onset of the perturbative regime. However it has another interesting effect: it changes the integrand in the cross section formulas, making it less singular at low values of $p_T$. Therefore, it
is expected that in CGC based models the separation between two regimes becomes less sensitive to the  choice of the separating cutoff. 

One can try to disentangle soft and semihard physics using experimental data. In \cite{biro} the authors performed a thorough analysis 
of the $p_T$ spectra measured in high energy proton-proton collisions, correlating them with the corresponding rapidity densities. 
They found that there are common features shared by all observables in hadronic collisions. This would be a manifestation of the dominance 
of soft interactions. However at $p_T \approx 0.6$ GeV significant deviations of the common trend start to appear. The authors attribute
this change to semihard interactions. Although this number may seem too small and its connection to the inputs of real calculations may 
seem unclear, it is an indication that the transition to the semihard regime may start at transverse momenta of the order of 1 GeV, rather 
than 5 or 10 GeV.  Other analyzes such as \cite{feal} and \cite{rosales} address the separation between soft and semihard physics using experimental data. In these works, the authors derived connections between the multiplicity and transverse momentum distributions and the separation scale was related to the effective temperature, which can be inferred from experimental data.

From a more phenomenological point of view, the attempts to understand data with two-component models was extremely successful. 
In the case of multiplicity distributions, a modified version of the study presented in \cite{halzen2} was proposed by 
Giovannini and Ugocioni \cite{giova1,giova3,giova2} and further developed in \cite{ghosh}. In these works the authors performed fits of the available multiplicity distributions with a double negative binomial distribution (DNBD). This kind of fit was also adopted by the experimental collaborations when presenting their results, as in \cite{alice1,alice2,alice3}. In all these works the individual distributions should come from one type of mechanism or source. It should be mentioned that the double source approach could be further extended to three sources and, consequently, to the use of three negative binomial distributions (TNBD). Indeed, it was shown in 
\cite{zbo} that in the TNBD approach it is possible to obtain excellent fits of the data and, in spite of larger 
number of parameters, still get insights on the data, specially on the low multiplicity region. 

Although the motivation for the DNBD was the separation between soft and semihard components, there was nothing in the used formulas connecting them to QCD. Very recently \cite{nos25} we performed a new DNBD fit,  but this time calculating the soft and semihard mean multiplicities with the $k_T$ factorization formula. In our formalism we not only have a clear connection to QCD but also this connection introduces new constraints and reduces the number of free parameters.  In our work we had to use an energy scale $\Lambda$ 
(which is $p_{Tmin}$) to separate soft and semihard processes. We used one single representative value which, in view of the discussion 
presented above,  was chosen to be $\Lambda = 1.4$ GeV. In the present work we repeat the procedure and the fits, changing the value of 
$\Lambda$ in order to test the sensitivity  of our results to this cutoff. 


\section[A two-component model with kT factorization]{A two-component model with $k_T$ factorization}

In the $k_T$ factorization approach the inclusive production cross-section is given by \cite{nos25,kov}:
\begin{equation}
\label{Kt-energia}
    E \frac{d \sigma} {d^3 p}= K \frac{4 \pi N_{c}} {N_{c}^{2}-1} \alpha_{s}(Q^2)\; \frac{1} {p_{{\perp}}^{2}} \;  
    \int^{p_{{\perp}}} \, d k_{{\perp}}^{2} \; \varphi_{1} ( x_{1}, k_{{\perp}}^{2} ) \; \varphi_{2} ( x_{2}, ( p-k )_{{\perp}}^{2} ),
\end{equation}
where $x_{1,2} = \frac{p_{\perp}}{\sqrt{s}}\exp{(\pm y)}$, $\sqrt{s}$ is the center-of-mass energy, $N_c$ is the number of colors and $K$ is a normalization factor that describes the conversion of partons to hadrons. $\varphi$ is the  unintegrated gluon distribution (UGD) of a proton, and it can be related to the gluon density by
\begin{equation}
\label{int-gluon}
    xG(x,\mu^2) = \int^{\mu^2} d k_{{\perp}}^{2} \varphi ( x, k_{{\perp}}^{2} ).
\end{equation}
Integrating (\ref{Kt-energia}) over $p_{\perp}$ we obtain the multiplicity per unit of rapidity:
\begin{equation}
    \frac{dN}{dy} = \frac{1}{S} \int_{p_{\perp{min}}}^{p_{\perp{max}}} d^2 p_{\perp}  E \frac{d \sigma} {d^3 p},
    \label{sep} 
\end{equation} 
where $S$ is a typical interaction area, and the minimum and maximum values of $p_{\perp}$ are determined by the experimental conditions. Using (\ref{int-gluon})  and the approximation described in
\cite{nos25}, we can rewrite (\ref{sep})  as:
\begin{align}
   \frac{dN}{dy} &= \frac{K}{S} \frac{4 \pi N_{c}} {N_{c}^{2}-1} \; \int_{p_{\perp{min}}}^{p_{\perp{max}}} \frac{d p_{\perp}^2} {p_{{\perp}}^{4}} \alpha_s(Q^2) xG_2(x_2, p_{\perp}^2)xG_1(x_1, p_{\perp}^2),
    \label{kt-final} 
\end{align} 
The multiplicative constants $K$ and $S$ can be grouped into a single parameter, which is fixed by fitting the available pseudorapidity distributions. For the UGD, we adopt the one from the GBW model (see \cite{nos25} for 
details and discussion). For the purposes of our work, the GBW model is especially appropriate because it reproduces very well the DIS data on $F_2$ at very low scales, which are consistent with the low transverse momenta of the 
bulk of produced particles at the LHC. 

In expression  Eq. (\ref{Kt-energia}),   and also in  Eq. (\ref{kt-final}), one gluon with $p_{\perp}$ is produced from the fusion of a gluon with $k_{\perp}$ and another gluon with
$p_{\perp} - k_{\perp}$. Therefore the momentum $p_{\perp}$ is, to a good approximation, the minimum value of the momentum flowing in the three gluon vertex, i.e. the one to be used in the running coupling constant. For $p_{\perp}$ larger than a few GeV, the coupling $\alpha_s(Q^2) $ 
becomes sufficiently smaller than one and we are in the perturbative domain. Following the reasoning presented in the introduction we  call these events semihard. The events with lower values of $p_{\perp}$  will be called soft. The total multiplicity per unit of rapidity is 
the sum of the contributions from  the soft and semihard events. The separation is achieved by introducing a cutoff $\Lambda$ in the integral over $p_{\perp}$, which defines the two contributions:
\begin{align}
\label{sep_final}
\frac{dN}{dy}
&= \frac{K}{S}\,\frac{4\pi N_c}{N_c^2-1}\Bigg[
    \int_{p_{\perp,\min}}^{\Lambda}\!\frac{dp_{\perp}^2}{p_{\perp}^4}\,\alpha_s(Q^2)\,xG_2(p_{\perp}^2)\,xG_1(p_{\perp}^2) \nonumber\\
&\qquad\qquad\qquad\quad
    +\int_{\Lambda}^{p_{\perp,\max}}\!\frac{dp_{\perp}^2}{p_{\perp}^4}\,\alpha_s(Q^2)\,xG_2(p_{\perp}^2)\,xG_1(p_{\perp}^2)
\Bigg]\nonumber\\
&= \frac{dN_s}{dy} + \frac{dN_{sh}}{dy}\,.
\end{align}
where $ N_s $ and $N_{sh}$ are the soft and semihard components respectively. 
Integrating (\ref{sep_final}) over the pseudorapidity  we obtain the average multiplicity:
\begin{equation}
  N  \, = \,   N_s  \, + \,  N_{sh}.
\label{n-cgc}
\end{equation} 
The only free parameter is the constant $K/S$ and it was determined in \cite{nos25} 
by using Eq. (\ref{sep_final}) to fit the measured pseudorapidity distributions.  In this fitting procedure we 
identified the number of produced gluons with the number of measured charged particles ("parton-hadron duality"). 
As discussed in \cite{nos25}, for our purposes this is a good approximation. 

At this point a remark is in order. Strictly speaking, this formalism is only valid for high energy scales and belongs to the perturbative domain of QCD. In Eq. (\ref{sep_final}) the rapidity distribution depends on the functions $G$, which in turn are obtained from the unintegrated gluon distributions  $\varphi$, which should be solutions of QCD evolution equations derived in the perturbative regime. In practice, since the evolution equations are too complicated, most authors use a QCD-inspired Ansatz for $\varphi$, which contains parameters. These parameters are fixed by fitting data on deep inelastic scattering. Part of these data is taken at very low $Q^2$ and certainly contain strong non-perturbative effects. Because of this procedure, the obtained UGDs are appropriate to describe other data which include non-perturbative physics, such as transverse momentum distributions of particles measured in minimum bias events. Examples of a good description of data can be found in \cite{andre1} and \cite{andre2}. In 
short  Eq. (\ref{sep_final}) was derived 
to study data dominated by perturbative physics (our semihard component). When it is extrapolated to lower energy scales (our soft component) it becomes a {\it model}, which should give good results because its parameters were 
fixed by fitting other data in the soft physics domain. 

\section[Double negative binomial fits and the dependence on Lambda]{Double negative binomial fits and the dependence on $\Lambda$} 

The DNBD is expressed as follows:
\begin{equation} 
\label{DNBD}
    P(n) = \lambda \; [ \; \alpha \; P(n,\langle n\rangle_s,k_s) + (1-\alpha) \; P(n,\langle n \rangle_{sh},k_{sh}) \; ],
\end{equation}
where
\begin{equation}
P(n,\langle n \rangle,k) = \frac{\Gamma(k+n)}{\Gamma(k) \, \Gamma(n+1)} \frac{\langle n \rangle^n  \, k^k}{(\langle n \rangle + k)^{n+k}}
\label{NB}.
\end{equation}
The parameter $\alpha$ represents the fraction of soft events; consequently, $(1-\alpha)$ denotes the fraction of semihard events;  $\langle n \rangle_s$ and $\langle n \rangle_{sh}$ 
are the mean multiplicities of soft and semihard events respectively; $k_s$ and $k_{sh}$ are the negative binomial parameters of the
soft and semihard components.  
The first few bins cannot be reproduced by any NBD and they are related to a different production mechanism, not included in our formalism. Therefore, these bins were removed from the fitting process. To account for this, a normalization factor $\lambda$ was introduced  in  \cite{alice1}. 
The average multiplicity obtained from  (\ref{DNBD}) can be expressed as:
\begin{equation}
    \langle n \rangle = \lambda\;[\;\alpha \;\langle n \rangle_s + (1-\alpha) \;\langle n \rangle_{sh}\;].
 \label{2nbd}
 \end{equation}
Comparing expression (\ref{2nbd}) with (\ref{n-cgc}), and recalling that $N = \langle n \rangle $, we conclude that:
 \begin{eqnarray}
\lambda \, \alpha  \, \langle n \rangle_s     &=& N_s , \nonumber \\  
\lambda \, (1-\alpha) \langle n \rangle_{sh}  &=& N_{sh}.  
\label{sis}
 \end{eqnarray}
Since $N_s$ and $N_{sh}$ are known, these two equations reduce the number of free parameters from six to four: $\alpha$, $\lambda$, $k_s$,
and $k_{sh}$.

\section[Results and discussion]{Results and discussion}

In \cite{nos25} we have used Eq. (\ref{DNBD}) to fit all the available multiplicity distributions measured at the 
LHC. The fit was of very good quality and we were able to determine the energy and rapidity window dependence 
of the main parameters. Before presenting new material, we will now review the findings of \cite{nos25}.  First we used Eq.(\ref{kt-final} )
to fit the measured pseudorapidity distributions, thereby fixing the parameter $K/S$, which turned out to be, within the uncertainties, independent of the energy. This  behavior is reasonable, given the interpretation of $K$ (as a  correction factor accounting for the parton-hadron conversion) and of $S$ (as a typical hadron-hadron interaction area). The next step was to perform a DNBD fit of the available multiplicity distributions, determining the values of $\lambda$, $\alpha$, $k_s$ and $k_{sh}$. Having obtained these quantities, we used (\ref{sis}) to 
determine $\langle n\rangle_s$ and $\langle n\rangle_{sh}$. 
As can be seen in Fig. \ref{fig2} and also in Figs. presented in \cite{nos25},  the dependence of the fitting parameters with the energy is smooth and the points could be fitted by some simple function, allowing us to make some predictions for higher energies. In the following subsections, we briefly comment on the results of our DNBD fits and compare them with other DNBD fits.

\subsection[lambda]{$\lambda$}

We found that, within the uncertainties, $\lambda$ is very close to one for all energies and rapidity windows. 
This indicates that the removal of the first bins did not strongly change the normalization of the multiplicity distributions.  A similar result was found in \cite{alice1}.  

\subsection[alpha]{$\alpha$}

The behavior of $\alpha$ is shown below in Fig. \ref{fig2}. It decreases with the energy for all rapidity windows. This is the most important conclusion of \cite{nos25} and is corroborated in the present study, as discussed below. In contrast, in the DNBD fit made in \cite{alice1}, the opposite is observed: the value of $\alpha$ tends to increase with energy. Moreover, in the present work, we find that $\alpha$ remains constant, within uncertainties, when varying the pseudorapidity cut. In contrast, the parameter $\alpha$ obtained in \cite{alice1} increases as wider pseudorapidity windows are considered. 


\subsection[ks and ksh]{$k_s$ and $k_{sh}$}

Our results show that the values of the parameters $k_s$ and $k_{sh}$ decrease with the collision energy in all rapidity windows, except  $|\eta| <1$, in which case they stay constant with the energy. This feature and the fact that in the narrower windows we observe KNO scaling and not in the others strongly suggest that there is a change 
of dynamics when we go from one region to the other. It is a challenge for theorists to understand this change.  

For comparison, the parameter $k_s$  found in \cite{alice1} decreases with the energy  for all  pseudorapidity windows.  The behavior of $k_{sh}$ in \cite{alice1} is difficult to determine and less consistent.

Fixing the energy and analyzing the variation with the rapidity window, in \cite{nos25} $k_s$ increases when the pseudorapidity window widens, while in \cite{alice1}  we observe a decrease. On the other hand, in both works 
$k_{sh}$ increases as the pseudorapidity window widens.

\subsection[ns and nsh]{$\langle n\rangle_s$ and $\langle n\rangle_{sh}$} 

The parameters $\langle n\rangle_s$ and $\langle n\rangle_{sh}$ follow the same pattern in both analyses, that is, they increase with energy and with the pseudorapidity window. 

The main advantage of our approach compared to the previous DNBD fits is that we obtain a more consistent and physical behavior for the parameters, especially for the parameter $\alpha$. In general, a fitting algorithm will always search for the best values to describe the data, which can lead to a non-physical behavior. We observe that by including the $k_T$ factorization for the determination of the DNBD parameters, we obtain a clear and more physical understanding of the proportion of soft and semihard events. Furthermore, our approach naturally leads to a reduction in the number of free parameters, and to a consistent behavior of $\alpha$ with energy across the analyzed pseudorapidity windows, favoring its predictability.

\subsection[Lambda]{$\Lambda$}
\begin{figure}[H]
\begin{adjustwidth}{-\extralength}{0cm}
  \centering
  \subfloat[]{%
    \includegraphics[width=0.44\textwidth]{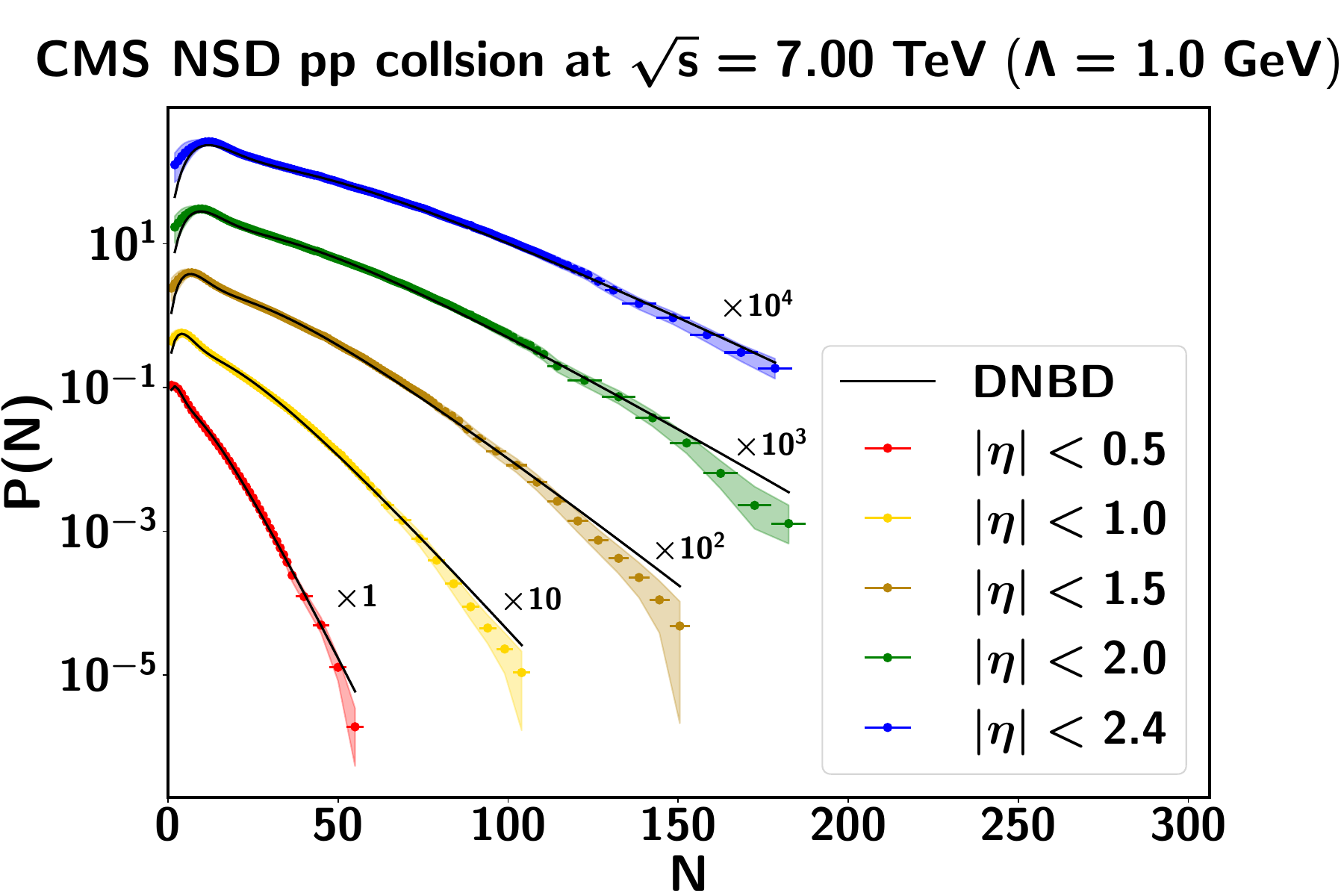}%
  }\hfill
  \subfloat[]{%
    \includegraphics[width=0.44\textwidth]{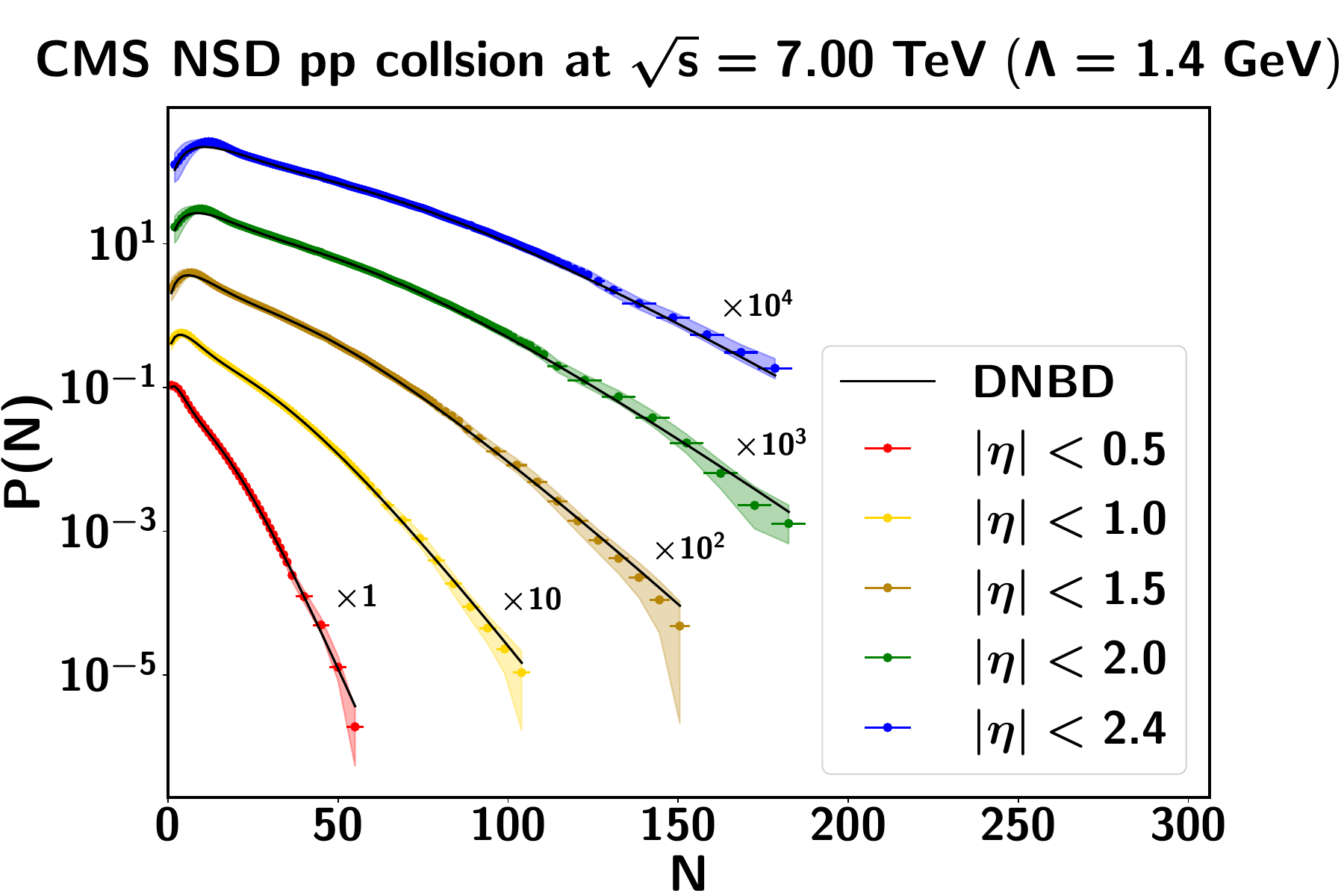}%
  }\hfill
  \subfloat[]{%
    \includegraphics[width=0.44\textwidth]{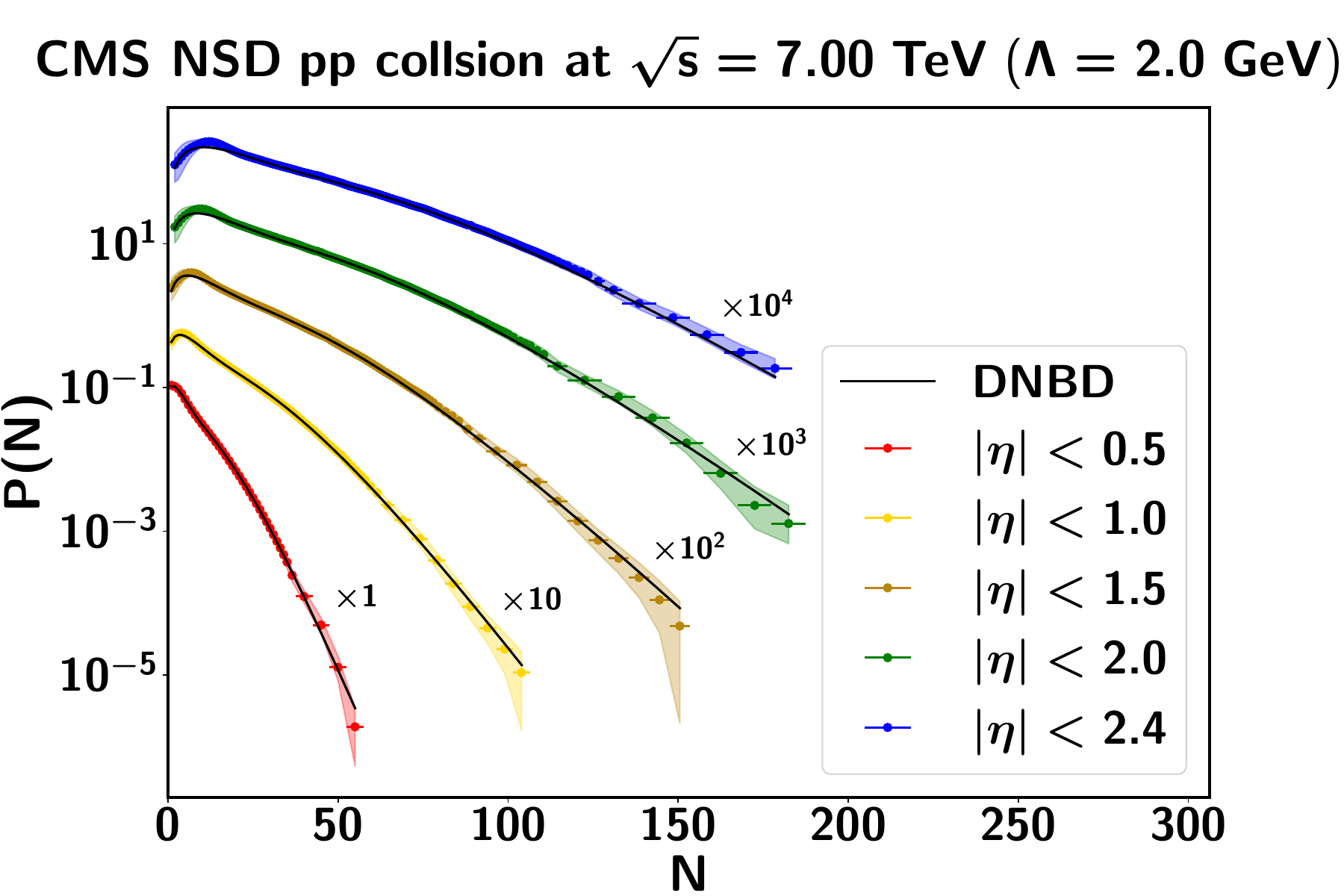}%
  }
\end{adjustwidth}
\caption[Multiplicity distributions measured in pp collisions at sqrts=7 TeV.]{Multiplicity distributions measured in pp collisions at $\sqrt{s} =7$ TeV. The data points are fitted with Eq. (\ref{DNBD}) taking fixed values of the cutoff $\Lambda$. (\textbf{a}) $\Lambda=1$ GeV. (\textbf{b}) $\Lambda=1.4$ GeV. (\textbf{c}) $\Lambda=2$ GeV.}
\label{fig1}
\end{figure}

In  \cite{nos25} we fixed the cutoff to be $\Lambda = 1.4$ GeV.
In view of the discussion in 
the introduction, this seems a reasonable value. However, to complete the discussion it is necessary to know how the 
results change when we change the value of $\Lambda$. In fact this is the main goal of the present work. It is 
straightforward to redo all the fits with different cutoff values. We took $\Lambda = 1$,~$1.4$ and $2$ GeV. The 
results are shown in Fig. \ref{fig1}. As can be seen, the three choices of $\Lambda$ lead to fits of similar quality. This statement can be made quantitative by computing the corresponding $\chi^2$, which is shown in Table \ref{tab1} and
which is small in all cases. We might be tempted to say that the lower cutoff is disfavored by data, but no firm 
conclusion be drawn yet.  For conciseness we only show the results obtained for $\sqrt{s} = 7$ TeV. We have checked that similar conclusions can be found studying the other energies.

\begin{figure}[H]
\begin{adjustwidth}{-\extralength}{0cm}
  \centering
  \subfloat[]{%
    \includegraphics[width=0.42\textwidth]{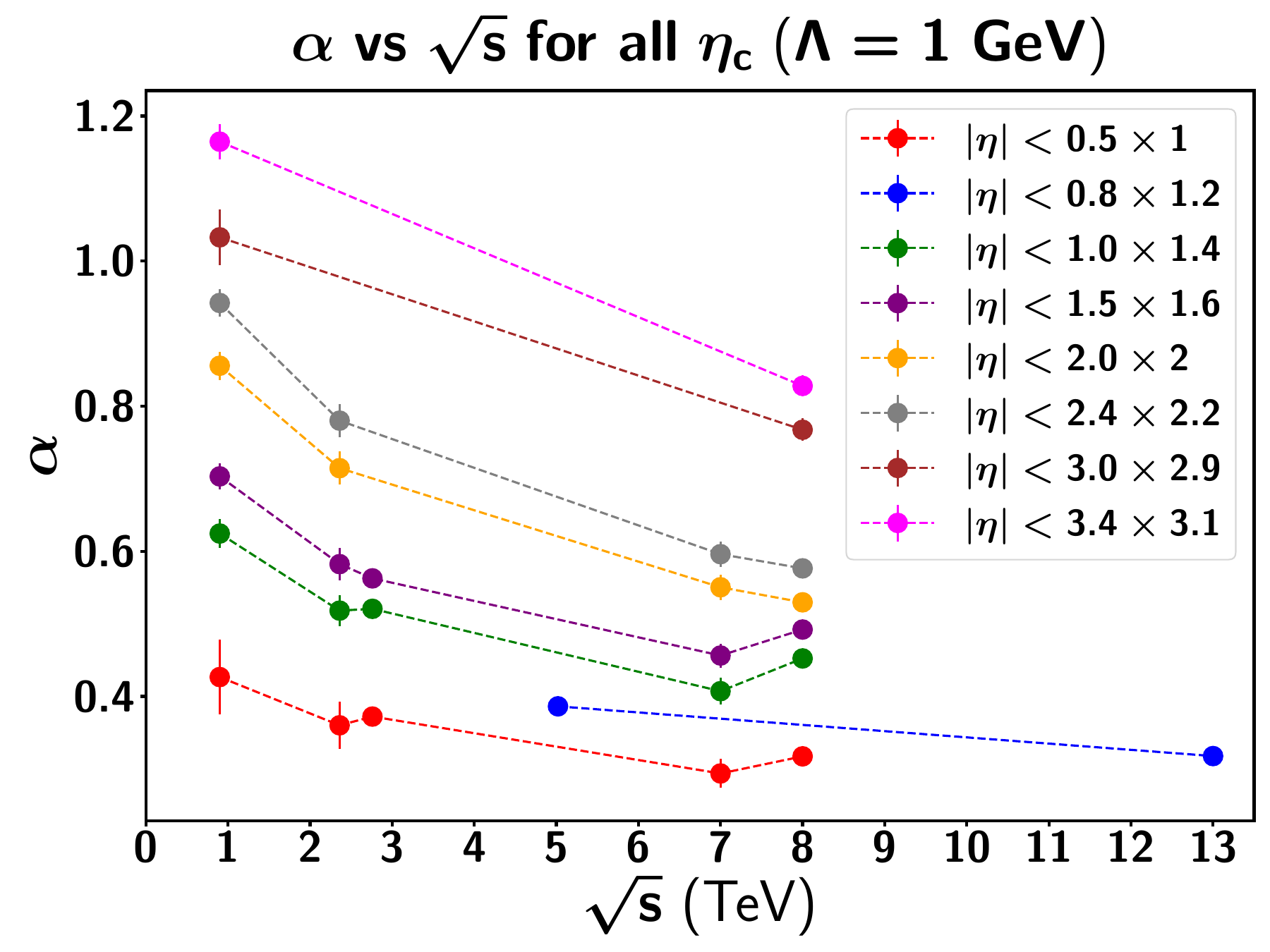}%
  }\hfill
  \subfloat[]{%
    \includegraphics[width=0.42\textwidth]{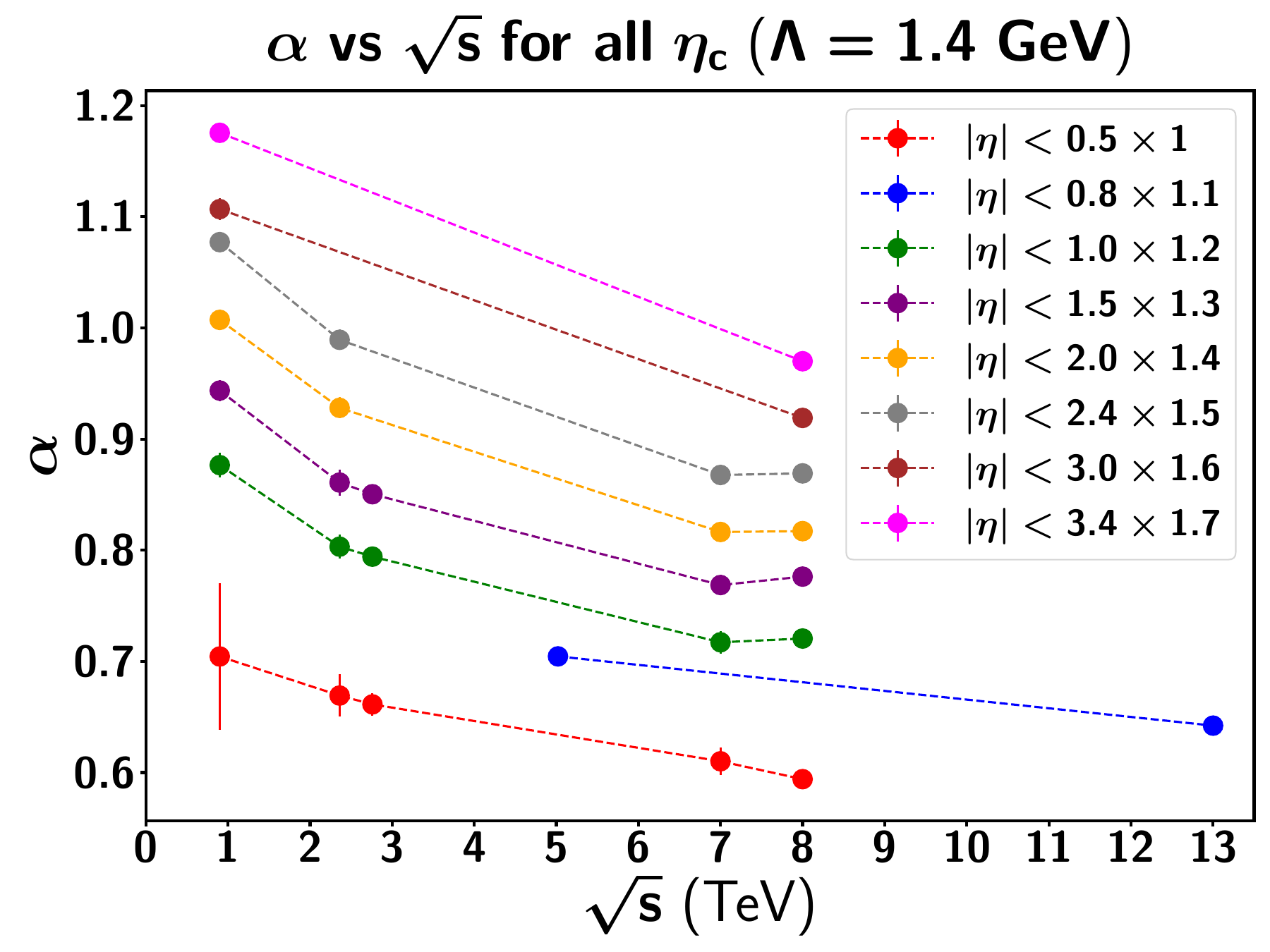}%
  }\hfill
  \subfloat[]{%
    \includegraphics[width=0.42\textwidth]{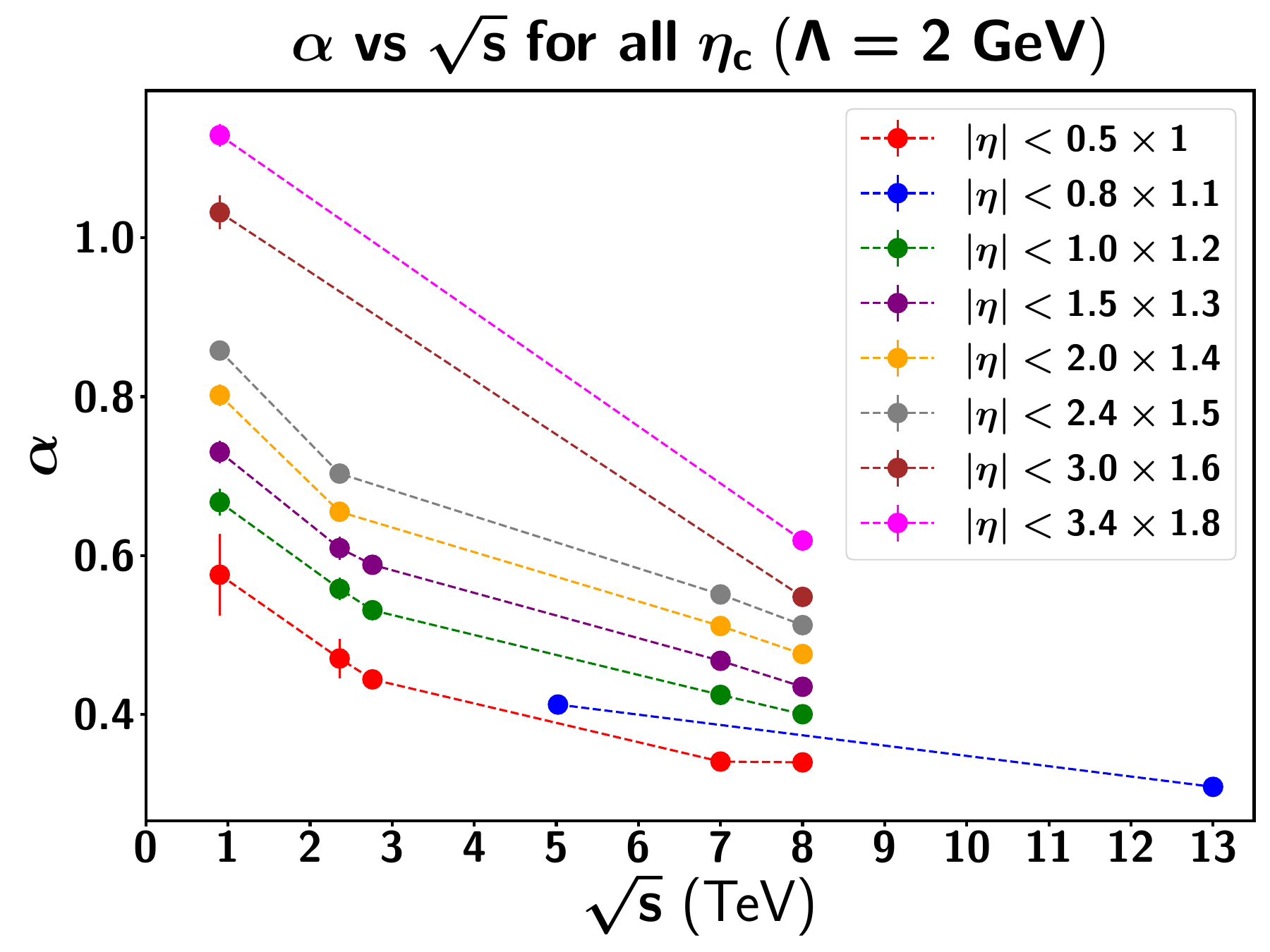}%
  }
\end{adjustwidth}
\caption[Parameter alpha as a function of the energy sqrts for different pseudorapidity windows.]{Parameter $\alpha$ as a function of the energy $\sqrt{s}$ for different pseudorapidity windows. The points are obtained from fitting the data and have been multiplied by different factors, for a better visualization. The lines are just to guide the eyes. In each panel the different curves refer to different pseudorapidity windows $-\eta_c < \eta < + \eta_c$. In each panel the value of the cutoff $\Lambda$ was fixed. (\textbf{a}) $\Lambda=1$ GeV. (\textbf{b}) $\Lambda=1.4$ GeV. (\textbf{c}) $\Lambda=2$ GeV.}
\label{fig2}
\end{figure}

In \cite{nos25} the most important conclusion was that the parameter $\alpha$ (the fraction of soft events), 
as expected, decreases with the energy $\sqrt{s}$ but, surprisingly, it decreases in the same way for all rapidity 
windows considered. In Fig. \ref{fig2} we can see that this decreasing trend remains valid for the three values of 
$\Lambda$.  In Fig.  \ref{fig2}c the two upper lines seem to fall faster than the others, but this may be due just to the
lack of points. From this analysis we conclude that the precise choice of the "frontier" between soft and semihard physics is not crucial and it is possible to fit the data with several values of the separating scale $\Lambda$. Moreover we confirm the previously found decreasing behavior of the parameter $\alpha$ with the energy. This 
relative insensitivity shows that the fitting procedure is robust.

\subsection[KNO scaling]{KNO scaling}

One interesting property of multiplicity distributions is the so-called KNO scaling, according to
which, at very high energies the multiplicity distribution $P(n)$, when multiplied by $\langle n \rangle$ and plotted as a function
of the variable $z = n/ \langle n \rangle$ becomes an energy-independent function. KNO scaling was shown to be valid for high energy jets \cite{dok,gui}, which suggests that it is a consequence of perturbative QCD. 
Ultimately, it would emerge asymptotically as a consequence of the cascading nature of parton multiplication.
However, KNO scaling was also observed in other experimental scenarios dominated by non-perturbative physics, such as minimum bias events, where it has a non-trivial behavior. Indeed, it has been observed that KNO scaling holds for central and narrow pseudorapidity windows, $|\eta| < 0.5$ and is clearly violated for $|\eta| > 1.0$.   

\section[Conclusion]{Conclusion} 

In a previous work \cite{nos25}, 
using a weighted superposition of two Negative Binomial Distributions Eq. (\ref{DNBD}) we fitted all the available multiplicity distributions measured at the LHC. The fit was of very good quality and we were able to determine the energy and rapidity window dependence of the main parameters. The separation between soft and semihard processes was
made through the introduction of an energy scale $\Lambda$. In \cite{nos25} it was chosen to be $\Lambda=1.4$ GeV and kept fixed. In this work we have completed our study, considering different values of $\Lambda$. The behavior of the 
main parameters with the energy and rapidity window remains the same. 
From this analysis we conclude that the precise choice of the "frontier" between soft and semihard physics is not crucial and it is possible to fit the data with several values of the separating scale $\Lambda$. Moreover, 
our analysis  suggests that there is no correlation between the dominance of soft (or semihard) processes and KNO scaling. The rapidity dependence of the KNO violations suggests that the central and forward pseudorapidity regions are governed by different aspects of QCD, probably related to the composition of the central source (mostly gluons) and the source in the fragmentation region (mostly valence quark remnants), as pointed out  long ago \cite{fowler}. Alternatively, the pseudorapidity window dependence of KNO scaling may be connected with the multiplicity scaling proposed in \cite{simak}. Further studies will be welcome.




\begin{table}[H]
\caption[Parameters fitted to adjust Eq. (DNBD) to the multiplicity distributions.]{Parameters fitted to adjust Eq. (\ref{DNBD}) to the multiplicity distributions measured in pp collisions at $\sqrt{s} = 7$ TeV.}
\label{tab1}
\begin{adjustwidth}{-\extralength}{0cm}
\begin{tabularx}{\fulllength}{CCCCCC}
\toprule
\textbf{$\Lambda$ (GeV)} & \textbf{$\eta_c$} & \textbf{$\alpha$} & \textbf{$\langle n \rangle_s$}
& \textbf{$\langle n \rangle_{sh}$} & \textbf{$\chi^2 / \mbox{dof}$} \\
\midrule
\multirow{5}{*}{1}	& 0.5 & 0.294 $ \pm $ 0.020	& 2.390 $ \pm $ 0.227 & 8.448 $ \pm $ 0.474 & 0.502\\
& 1.0	& 0.291 $ \pm $ 0.013 & 4.912 $ \pm $ 0.318 & 16.879 $\pm $ 0.636 & 0.495\\
& 1.5 & 0.285 $ \pm $ 0.010 & 7.706 $ \pm $ 0.406 & 25.415 $\pm $ 0.764 & 0.568\\
& 2.0 & 0.275 $ \pm $ 0.009 & 10.956 $\pm $ 0.501 & 34.049 $\pm $ 0.859 & 0.492\\
& 2.4 & 0.271 $ \pm $ 0.008 & 13.473 $\pm $ 0.572 & 40.700 $\pm $ 0.936 & 0.442\\
\midrule
\multirow{5}{*}{1.4} & 0.5 & 0.610 $ \pm $ 0.012	& 3.376 $ \pm $ 0.182 & 11.214 $\pm $ 0.711 & 0.102\\
& 1.0	& 0.598 $ \pm $ 0.008 & 6.994 $ \pm $ 0.267 & 21.978 $\pm $ 0.943 & 0.113\\
& 1.5 & 0.591 $ \pm $ 0.007 & 10.723 $\pm $ 0.347 & 32.745 $\pm $ 1.153 & 0.188\\
& 2.0 & 0.583 $ \pm $ 0.006 & 14.731 $\pm $ 0.447 & 43.493 $\pm $ 1.400 & 0.310\\
& 2.4 & 0.578 $ \pm $ 0.006 & 17.846 $\pm $ 0.504 & 51.731 $\pm $ 1.520 & 0.411\\
\midrule
\multirow{5}{*}{2.0} & 0.5 & 0.341 $ \pm $ 0.011	& 11.798 $ \pm $ 0.299 & 3.601 $ \pm $ 0.117 & 0.093\\
& 1.0	& 0.354 $ \pm $ 0.007 & 23.035 $ \pm $ 0.452 & 7.469 $ \pm $ 0.200 & 0.106\\
& 1.5 & 0.360 $ \pm $ 0.006 & 34.274 $ \pm $ 0.601 & 11.456 $ \pm $ 0.287 & 0.182\\
& 2.0 & 0.365 $ \pm $ 0.006 & 45.537 $ \pm $ 0.776 & 15.733 $ \pm $ 0.386 & 0.313\\
& 2.4 & 0.367 $ \pm $ 0.005 & 54.243 $ \pm $ 0.911 & 19.093 $ \pm $ 0.456 & 0.420\\
\bottomrule
\end{tabularx}
\end{adjustwidth}
\end{table}

\authorcontributions{Conceptualization, H.M.F. and F.S.N.; Methodology, H.M.F. and F.S.N.; 
Software, H.M.F.; 
Writing---review and editing, H.M.F. and F.S.N.; 
Supervision, F.S.N.;  
Funding acquisition, H.M.F. and F.S.N. 
All authors have read and agreed to the published version of the manuscript.}

\funding{ This work was partially financed by Funda\c{c}\~ao de Amparo \`a Pesquisa do Estado de S\~ao 
Paulo (FAPESP), grant no. 2024/13426-0, Conselho Nacional de Desenvolvimento Cient\'{\i}fico e Tecnol\'ogico 
(CNPq), grant no. 309262/2019-4 and by the INCT-FNA.}

\acknowledgments{We are grateful to A. V. Giannini, M. V. T. Machado, M. Munhoz and Y. Lima for useful discussions.}

\conflictsofinterest{The authors declare no conflicts of interest. The funders had no role in the design
of the study; in the collection, analysis or interpretation of data; in the writing of the manuscript, or
in the decision to publish the results.}


\begin{adjustwidth}{-\extralength}{0cm}

\reftitle{References}





\isChicagoStyle{%

}{}

\isAPAStyle{%

}{}


%


\PublishersNote{}
\end{adjustwidth}
\end{document}